\documentclass[runningheads]{llncs}
\usepackage[T1]{fontenc}
\usepackage{graphicx}
\usepackage{tikz}
\usepackage{xcolor}
\usepackage[ruled,vlined, linesnumbered]{algorithm2e}
\usepackage{xspace}
\usepackage{amssymb}
\usepackage{amsmath}
\usepackage{placeins}
\usepackage{subcaption}
\usepackage{url}
\usepackage{array}
\usepackage{listings}
\lstdefinestyle{lean}{basicstyle=\ttfamily\small, columns=fullflexible, keepspaces=true, keywordstyle=\bfseries,
  morekeywords={def,theorem,inductive,Prop,List,fun,match,Type},
  literate={∀}{{$\forall$}}1 {ℕ}{{$\mathbb{N}$}}1 {→}{{$\to$}}1 {≤}{{$\le$}}1 {ℤ}{{$\mathbb{Z}$}}1 {∃}{{$\exists$}}1 {∧}{{$\wedge$}}1 {∈}{{$\in$}}1}

\definecolor{safeA1}{RGB}{26,133,255}  
\definecolor{safeA2}{RGB}{212,17,89}   

\newcommand{\lean}[1]{\texttt{\detokenize{#1}}}

\newcommand{\lowerbound}{22}
\newcommand{\ifoptimal}[1]{}

\newcommand{\Ftwo}{\ensuremath{\mathbb{F}_2}\xspace}

\begin{document}
\title{Lower Bound of 22 for $3\times3$ Matrix Multiplication over $\mathbb{Z}$}
%
%
\author{Isaac Rudich\inst{1, 2}\orcidID{0000-0002-3106-1020} \and
Louis-Martin Rousseau\inst{1}\orcidID{0000-0001-6949-6014}}
\authorrunning{I. Rudich and L.M. Rousseau}
\institute{Polytechnique Montréal, Quebec, Canada \and
Carnegie Mellon University, Pittsburgh PA, USA}
%
\maketitle              
\begin{abstract}
Strassen showed that two $2\times2$ matrices can be multiplied with 7 multiplications instead of 8. 
Applied recursively, his algorithm multiplies two $n\times n$ matrices with $\approx O(n^{2.807})$ multiplications, beating the naive $O(n^{3})$.
The best known $3\times3$ recursive matrix multiplication algorithm uses 23 multiplications ($\approx O(n^{2.854})$). The best published lower bound of 21 (on algorithms with integer constants) leaves room for an algorithm with $\approx O(n^{2.771})$ mutiplications, and thus does not rule out the possibility of an algorithm that would beats Strassen's. 

We prove a lower bound of 22 multiplications for any $3\times3$ recursive algorithm with integer constants, proving that no such algorithm can do better than $\approx O(n^{2.814})$ multiplications, and eliminating the possibility of a $3\times3$ algorithm that beats Strassen's $2\times2$ method. The proof builds on a recent decomposition method from Wang, who approached the problem by turning it into 496 subproblems. We provide exact solutions for 359 of them. 
The proof is in Lean; verification requires auditing only a few short files. 
The Lean formalization directly encodes statements about the limitations of recursive algorithms for matrix multiplication, as opposed to just a statement about the rank of the problem. 
\end{abstract}

\section{Introduction}
\label{sec:intro}

Naively multiplying two $n\times n$ matrices performs $n^3$ multiplications of numbers.
Strassen showed that two $2\times2$ matrices can be multiplied with 7 multiplications instead of 8~\cite{strassen1969gaussian}. 
He also showed that his algorithm can be applied recursively: an $n\times n$ matrix can be split into four $\tfrac{n}{2}\times\tfrac{n}{2}$ blocks, with each block treated as a single entry of a $2\times2$ matrix. 
Strassen's algorithm uses $\mathcal{O}(n^{\log_2 7})$ ($\log_2 7 \approx 2.807$) multiplications, and was proven optimal for $2\times2$ matrices in 1971~\cite{hopcroft1969minimizing,hopcroft1971minimizing,winograd1971multiplication}.

The same idea works for any fixed size: an algorithm that multiplies two $3\times3$ matrices with $r$ multiplications, and works on blocks, multiplies two $n\times n$ matrices with $\mathcal{O}(n^{\log_3 r})$ multiplications.
Therefore a $3\times3$ algorithm improves on Strassen's exponent exactly when $r \leq 21$ ($\log_3 21 \approx 2.771$) and is worse for $r \geq 22$ ($\log_3 22 \approx 2.814$)~\cite{blaser2013fast}.
Laderman gave an algorithm for $3\times3$ that uses 23 multiplications and works on blocks~\cite{laderman1976noncommutative}, but there have not been any improvements on this since 1976.

Most lower bounds on the number of multiplications required are proven for a restricted class of algorithms, called \emph{bilinear algorithms}. 
Bilinear algorithms obtain a list of products by multiplying a linear combination of entries from the first matrix by a linear combination of entries from the second. 
Then they obtain each entry of the output matrix from a linear combination of those products.
The fewest products required by a bilinear algorithm for a problem is the \emph{rank} of the problem.
Strassen's and Laderman's algorithms are both bilinear, so the rank of $3\times3$ matrix multiplication is at most 23.
The rank can depend on which numbers are allowed to define the linear combinations.
For example, the constants may be restricted to \Ftwo, the field whose only elements are 0 and 1, and in which $1+1=0$.

In 2003, Bläser proved that the rank of $3\times3$ matrix multiplication is at least 19~\cite{blaser2003complexity,blaser2013fast}.
In March 2026, Wang built on the \emph{substitution method}~\cite{pan1966methods,landsberg2017abelian,hopcroft1969minimizing} and raised this to 20 over \Ftwo with an automated proof that was independently verified~\cite{wang2026automated,beuchert2026independent}.
He sorted every way of imposing certain condition types on $3\times3$ matrix multiplication into 496 problem classes, a list we call \emph{Wang's table}, and bounded the rank of each.
In September 2026, within ten days of each other, Wang~\cite{wang2026lower21} and Yang et al.~\cite{yang2026structural} independently proved that the rank over \Ftwo is at least 21, and Tahir obtained 21 as a conditional bound in a public review draft~\cite{tahir2026certificate}.
Yang et al.\ also formalized their proof in Lean, a program that automatically verifies mathematical proofs~\cite{demoura2021lean4}.

We call a class in Wang's table \emph{exact} when its rank is known, meaning we have a lower bound on its rank and an upper bound that matches.
We achieved a proof that the rank of $3\times3$ matrix multiplication over \Ftwo is at least \lowerbound, proving that matrix multiplication algorithms with integer constants that recurse are more efficient on $2\times2$ than $3\times3$. We did this by first finding strong upper bounds for the classes in Wang's table by exploiting symmetry across table elements, and then using those upper bounds as a map for which lower bounds were worth pursuing. Our proof is in Lean and leverages the Mathlib library~\cite{mathlib2020}. To the best of our knowledge, no lower bound above 21 had been published as of September 29, 2026.

By Hopcroft and Kerr, a lower bound over \Ftwo also holds for algorithms with integer constants that work on blocks, since those cannot rely on commutativity~\cite{hopcroft1969minimizing,hopcroft1973duality,amitsur1950minimal}.
Our Lean proof includes this reduction, so its final claim is about algorithms, not just rank.

Our contributions are as follows:
(1) we prove a lower bound of \lowerbound ~multiplcations required for $3\times3$ matrix multiplication by any algorthm that works on blocks with integer constants;
(2) we prove in Lean the exact value of 359 of the 496 classes of Wang's table, and lower and upper bounds for the other 137;
(3) we build an easy-to-audit approximatley 1,000 line Lean library that defines algorithms for matrix multiplcation, defines Wang's table, and connects the table to statements about algorithms; it is possible to verify the entire 1,000,000-line proof without going beyond those 1,000 lines;
and (4) we expand on an existing construction~\cite{blaser2013fast,yang2026structural} to glue algorithms from two different table classes into an algorithm for a third, which gave us 145 of our upper bounds.
AI agents (Claude, by Anthropic) wrote the search code and the Lean proofs that do not require a human auditor, under our direction. The audit-surface was human designed.

\section{Background}
\label{sec:background}

\subsection{Decompositions}
\label{sec:background:decompositions}

Number the entries of the $3\times3$ matrices $A$, $B$ and $C = AB$ from 1 to 9, row by row, and let $T$ be the $9\times9\times9$ tensor with $T_{ijk} = 1$ when the product of entry $i$ of $A$ and entry $j$ of $B$ appears in entry $k$ of $C$, and $T_{ijk} = 0$ otherwise.
An algorithm with $r$ products is given by a decomposition of $T$ into three $9\times r$ matrices $U$, $V$ and $W$.
Product $\ell$ multiplies a combination of entries of $A$, weighted by column $\ell$ of $U$, by a combination of entries of $B$, weighted by column $\ell$ of $V$.
Entry $k$ of $C$ is the sum of the products weighted by row $k$ of $W$.
The algorithm is correct for matrix multiplication exactly when
\[ T_{ijk} = \sum_{\ell=1}^{r} U_{i\ell} V_{j\ell} W_{k\ell} \quad \text{for all } i, j, k. \]
For an in-depth worked example, see Deza et al.'s Figure~1~\cite{deza2023fast}.

A lower bound of $r$ over \Ftwo means that no decomposition with fewer than $r$ columns and entries in \Ftwo exists, which is a finite search problem.
Our project began as an attempt to create an exact solver for this problem inspired by the work of Deza et al.\ who search for decompositions with constraint programming~\cite{deza2023fast}, and Heule et al.\ who search with a SAT solver~\cite{heule2021new}.
Both find decompositions well, but proving that none exists is much harder.

\subsection{Wang's Table}
\label{sec:background:table}

Wang's method bounds the hard problem of identifying exact rank by solving many easier ones~\cite{wang2026automated}.
A \emph{condition} requires a set of entries of $A$ to add up to 0, and a set of conditions restricts the problem to the matrices $A$ that meet them.
Its rank is the fewest products needed when $A$ meets the conditions and $B$ is arbitrary.
Up to symmetry (multiplying $A$ by invertible matrices on either side, or transposing it), and counting two sets of conditions as one when each implies the other, there are 496 classes.
We arrange them in rows by their number of independent conditions, from row~9 ($A = 0$) to row~0 (matrix multiplication itself).

In linear-algebra terms, a condition is a nonzero vector $x \in \Ftwo^9$, and $A$ meets it when $\sum_i x_i A_i = 0$.
A set of $d$ independent conditions spans a $d$-dimensional subspace $S$ of $\Ftwo^9$, and two sets that span the same $S$ allow the same matrices $A$, so the problem depends only on $S$, which Wang calls a constraint subspace~\cite{wang2026automated}.
We write $R(S)$ for the rank of its problem.
The row of a class is $\dim S$: row~9 is $S = \Ftwo^9$, which allows only $A = 0$, and row~0 is $S = \{0\}$.
If $S \subseteq S'$ then $R(S') \le R(S)$, because an algorithm that is correct for every $A$ allowed by $S$ is also correct for the $A$ allowed by $S'$~\cite{wang2026automated}.

\section{What Is Claimed and How to Audit It}
\label{sec:design}
Lean is a programming language for writing mathematical proofs: a theorem compiles only if Lean's kernel checks every step of its proof, and Lean can list the axioms each proof relies on~\cite{demoura2021lean4}.
So if an auditor is convinced that the theorems are stated correctly, and the code compiles using only Lean's standard axioms, the proof is verified automatically.
In the code below, \lean{def} defines a function, a \lean{Prop} is a proposition (a statement that is either true or false),  and a \lean{theorem} states a proposition whose proof must compile.

Any algorithm that multiplies two $3\times3$ matrices needs at least \lowerbound ~multiplications, under the conditions that (1) its constants are integers, and (2) it can be applied recursively by dividing a larger matrix into a $3\times3$ grid of square blocks of any size.
In our Lean formalization, an algorithm (\textit{program} in the code) is represented by a list of simple steps. 
This is built on the definition \lean{DoesMatrixMultiplicationOnBlocks}, which is a \lean{Prop} that is true if and only if a program multiplies $3\times3$ grids of blocks correctly, for any size of blocks.
That definition and the terms that define it are covered in Section~\ref{sec:layer01}; here they show the idea we are working towards.

\begin{lstlisting}[style=lean]
def RecursiveAlgorithmMultiplicationsAtLeast (n : ℕ) : Prop :=
  ∀ steps : List Step, DoesMatrixMultiplicationOnBlocks steps →
    n ≤ multiplication_count steps
\end{lstlisting}

This definition takes one input, $n$, a natural number, and returns a \lean{Prop}.
The statement reads: for every program \lean{steps}, if the program does matrix multiplication on blocks, then the number of multiplications of two values it performs is at least $n$.
So \lean{RecursiveAlgorithmMultiplicationsAtLeast n} is true if and only if every program that does matrix multiplication on blocks uses at least $n$ multiplications.

\begin{equation}\label{thm:main}
\begin{minipage}{0.9\linewidth}
\begin{lstlisting}[style=lean]
theorem recursive_algorithm_multiplications_at_least_22 :
    RecursiveAlgorithmMultiplicationsAtLeast 22
\end{lstlisting}
\end{minipage}
\end{equation}

A theorem has a name and a statement after the colon.
Theorem~\ref{thm:main} says that every program that does matrix multiplication on blocks uses at least 22 multiplications.

The Lean proof is split into layers.
Layers 01 to 03 each have a \emph{Defs} file, which defines every term the layer uses, and a \emph{Theorems} file, which states the layer's claims in those terms; layer 04 has only a Theorems file.
Each proof in a Theorems file is a one-line pointer into the \emph{aux} files, which hold the actual proofs and sit in a separate folder that an auditor never needs to open.
Layer 01 defines programs and decompositions and connects them. Layer 02 defines Wang's table and connects it to programs. Layer 03's Defs file is just a one-line shortcut for table lookups. We designed all of the definitions, and the theorems of layers 01 and 02, before any of their proofs existed.
Layer 03's Theorems file states bounds for all 496 elements of the table, from row 9 to row 0, and layer 04 states only Theorem~\ref{thm:main}.

To check the proof, an auditor:
(1) reads all the Defs files and the Theorems files of layers 01, 02 and 04;
(2) builds the repository with Lean v4.30.0 and Mathlib v4.30.0~\cite{demoura2021lean4,mathlib2020}, which checks every proof;
(3) checks that the \texttt{\#print axioms} line under every claim lists only Lean's three standard axioms (\lean{propext}, \lean{Classical.choice} and \lean{Quot.sound}),
and (4) replays the build with an independent kernel checker, so that nothing in the aux files can change the result of step~3.
The repository \url{https://doi.org/10.5281/zenodo.23047901} accompanies the paper.

The proof reaches Theorem~\ref{thm:main} in four steps.
Layer 01 shows that a program that works on blocks yields a decomposition with integer entries and no more columns than the program has multiplications.
Layer 02 shows that such a decomposition, read modulo 2, is a decomposition over \Ftwo, so a lower bound for matrix multiplication over \Ftwo bounds every such program.
Layer 03 bounds the classes of Wang's table row by row, from row~9 down to row~0, where the bound of 22 comes from the rank-sum lemma of Yang et al.~\cite{yang2026structural} applied to the two classes of row~1 that need 21 products.
Layer 04 turns the bound for row~0 into Theorem~\ref{thm:main} through layer 02's reduction.

\section{Programs and Decompositions}
\label{sec:layer01}

\subsection{Programs}

Layer 01 builds programs from the following steps.

\begin{lstlisting}[style=lean]
inductive Step
    fromA (i j : Fin 3)      fromB (i j : Fin 3)
    add (x y : ℕ)            sub (x y : ℕ)
    scale (w : ℤ) (x : ℕ)    mul (x y : ℕ)
    toC (i j : Fin 3) (x : ℕ)
\end{lstlisting}

A \lean{Step} either reads entry $(i, j)$ of $A$ or $B$ and performs an operation, or stores the value of step $x$ in entry $(i, j)$ of the output.
How each step behaves is fixed by \lean{step_value}.

\begin{lstlisting}[style=lean]
def step_value {R : Type} [Ring R] (s : Step) (A B : Matrix (Fin 3) (Fin 3) R)
    (value_at : List R) : R :=
  match s with
  | .fromA i j => A i j
  | .fromB i j => B i j
  | .add x y => value_at[x]?.getD 0 + value_at[y]?.getD 0
  | .sub x y => value_at[x]?.getD 0 - value_at[y]?.getD 0
  | .scale w x => (w : R) * value_at[x]?.getD 0
  | .mul x y => value_at[x]?.getD 0 * value_at[y]?.getD 0
  | .toC _ _ x => value_at[x]?.getD 0
\end{lstlisting}

The inputs of \lean{step value} are a step, the input matrices $A$ and $B$, and the list of values of the steps already run (\lean{value_at}).
Each line of the \lean{match} handles one kind of step, for example \lean{add} adds the values of steps $x$ and $y$.
Earlier values are looked up with \lean{value_at[x]?.getD 0}, which gives 0 if step $x$ has not run yet, so any list of steps is well defined.
All values lie in a type $R$, and \lean{[Ring R]} requires that $R$ has addition, subtraction and multiplication, which is all \lean{step_value} uses.
Real numbers qualify, and so do $m\times m$ real matrices, so the same program can be run with blocks as its values.
Whether a program returns the correct product when its values are blocks is what \lean{DoesMatrixMultiplicationOnBlocks} asks.
A program's cost, \lean{multiplication_count}, is its number of \lean{mul} steps.
Running a program on inputs $A$ and $B$, \lean{program_output} computes every step in order with \lean{step_value} and returns the $3\times3$ matrix whose entry $(i, j)$ is the value of the last \lean{toC i j} step, or 0 if there is none.

\begin{lstlisting}[style=lean]
def DoesMatrixMultiplicationOnBlocks (steps : List Step) : Prop :=
  ∀ m : ℕ, ∀ A B : BlockGrid m, program_output steps A B = A * B
\end{lstlisting}

This definition takes one input, \lean{steps}, which is a program given as a list of steps, and returns a \lean{Prop}, a statement that is either true or false.
Here \lean{BlockGrid m} is a $3\times3$ matrix whose entries are themselves $m\times m$ matrices of real numbers.
The statement after \lean{:=} reads: for every natural number $m$, and every $A$ and $B$ of type \lean{BlockGrid m}, running the program on $A$ and $B$ returns the product $A * B$.
So \lean{DoesMatrixMultiplicationOnBlocks steps} is true if and only if the program multiplies $3\times3$ grids of blocks correctly, whatever the size of the blocks.
\\\\
\noindent \textbf{Important Detail:} Because \lean{scale} takes a value in $\mathbb{Z}$, algorithms can only scale the values in $A$ and $B$ by integers. This is what restricts statements about this program to statements about algorithms with integer constants.

\subsubsection{Comparison with Hopcroft and Kerr's Model}
\label{sec:layer01:hk}
Our programs are close to the straight-line programs of Hopcroft and Kerr~\cite{hopcroft1969minimizing}, with these differences:
\begin{itemize}
\item their instructions may also add fixed integers, while ours use a constant only through \lean{scale}. Their added constants can be removed without increasing the number of multiplications~\cite{hopcroft1969minimizing}, but we did not formalize that step.
\item we judge correctness on real blocks of every size rather than as an identity in noncommuting variables. The two approaches are equivalent by Amitsur and Levitzki~\cite{amitsur1950minimal}.
\end{itemize}

\subsection{Decompositions}
\label{sec:layer01:decompositions}

In Lean, a decomposition is the object of Section~\ref{sec:background:decompositions}: a list of columns, each a triple of integer $3\times3$ matrices holding the matching columns of $U$, $V$ and $W$, with each column's 9 entries laid out as a $3\times3$ matrix.
It is valid when it computes $AB$ for every pair of real matrices $A$ and $B$ (\lean{IsValidDecomposition}). We define \lean{decomposition_to_program}, which takes in a decomposition and turns it into a program.
Layer 01 proves five facts about decompositions.

\textbf{(1)} We define \lean{DoesMatrixMultiplication} as a sanity check that is true exactly when a program performs matrix multiplication correctly for all real matrices. Then we prove that a decomposition is valid if and only if the program returned by \lean{decomposition_to_program} satisfies \lean{DoesMatrixMultiplication}. This establishes that our definition of \lean{IsValidDecomposition} is correct.
\textbf{(2)} \lean{decomposition_to_program} returns a program with a number of multiplications equal to the number of columns of the decomposition.
\textbf{(3)} Given a valid decomposition with $r$ columns, there exists a valid decomposition for $3^k\times3^k$ matrices with $r^k$ columns~\cite{strassen1969gaussian,pan1984how}. 

\textbf{(4)} A decomposition is valid if and only if the program returned by

\noindent \lean{decomposition_to_program} satisfies \lean{DoesMatrixMultiplicationOnBlocks}.
Together with (1), this means that for any decomposition, three statements are equivalent: the decomposition is valid, its program does matrix multiplication on real matrices, and its program does matrix multiplication on blocks.
So our three definitions agree wherever they can be compared, which is what shows that each of them is well-stated.
In particular, every valid decomposition yields a program that works on blocks, so the programs Theorem~\ref{thm:main} is about include every such algorithm (\lean{RecursiveAlgorithmMultiplicationsAtLeast}).

\textbf{(5)} For any program that works on blocks, there exists a decomposition that uses the same number of multiplications~\cite{hopcroft1969minimizing,hopcroft1973duality,amitsur1950minimal}. Without the block condition the fifth fact is false.
Winograd's inner-product algorithm~\cite{winograd1968inner} uses about half the multiplications of the naive method on large matrices, but each of its products mixes entries of $A$ and $B$, so it is not a decomposition, and it does not work on blocks.
For $3\times3$ matrices, Rosowski's commutative algorithm uses 21 multiplications~\cite{rosowski2019fast}, fewer than any decomposition, but it cannot be applied recursively.

\begin{equation}\label{thm:recursion}
\begin{minipage}{0.9\linewidth}
\begin{lstlisting}[style=lean]
theorem recursion_costs_a_decomposition (steps : List Step) :
    DoesMatrixMultiplicationOnBlocks steps →
      ∃ L : Decomposition, IsValidDecomposition L ∧
        L.length ≤ multiplication_count steps
\end{lstlisting}
\end{minipage}
\end{equation}

Theorem~\ref{thm:recursion} takes one input, \lean{steps}, a program, and reads: if the program works on blocks, some valid decomposition \lean{L} has at most as many columns as the program has multiplications.
With fact (3), a program with $r$ \lean{mul} steps that works on blocks therefore yields a valid decomposition for $3^k\times3^k$ matrices with at most $r^k$ columns, which is the recursion.
Theorem~\ref{thm:main} builds directly on Theorem~\ref{thm:recursion}: a lower bound on the columns of every valid decomposition is a lower bound on the multiplications of every program that works on blocks.

\section{Wang's Table}
\label{sec:layer02}

\subsection{Restricted Problems}

Layer 02 runs the programs of layer 01 with values in \Ftwo.
\Ftwo is a ring, so \lean{step_value} applies unchanged.
A condition is a $3\times3$ matrix $x$ over \Ftwo (a \lean{Bit3x3Matrix}), and $A$ meets it (\lean{MeetsCondition}) when the entries of $A$ at the cells where $x$ has a 1 add up to 0, as in Section~\ref{sec:background:table}.

\begin{lstlisting}[style=lean]
def IsValidTableDecomposition (conditions : List Bit3x3Matrix)
    (L : Decomposition) : Prop :=
  ∀ A B : Bit3x3Matrix, (∀ x ∈ conditions, MeetsCondition A x) →
    program_output (decomposition_to_program L) A B = A * B
\end{lstlisting}

This definition takes two inputs, a list of conditions and a decomposition, and returns a \lean{Prop}.
The statement reads: for all $3\times3$ matrices $A$ and $B$ over \Ftwo, if $A$ meets every condition, then the program of the decomposition, run on $A$ and $B$, returns $A * B$.
So \lean{IsValidTableDecomposition conditions L} is true if and only if the decomposition multiplies correctly over \Ftwo whenever $A$ meets the conditions.
\lean{TableAtLeast conditions n} says that every decomposition valid for the conditions has at least $n$ columns, \lean{TableAtMost conditions n} says that one has at most $n$, and \lean{TableExactly} says that both hold.

\subsection{Canonicalization of the Table}

Layer 02 generates Wang's table from its definition rather than copying it.
A condition is stored as a 9-bit number, a mask (\lean{mask_matrix}).
Every list of conditions implies the same conditions as exactly one reduced basis, a sorted list of masks in reduced echelon form (\lean{IsReducedBasis}).
Two lists are equivalent (\lean{Equivalent}) when one of the symmetries of Section~\ref{sec:background:table} maps the conditions implied by one exactly onto those implied by the other.
The table's element for each class is the reduced basis that comes first in dictionary order among those of the class (\lean{IsCanonical}), and the table lists these with more conditions first, then in dictionary order (\lean{table}).
Layer 02 proves the following facts about it.

\textbf{(1)} The table has 496 elements, and rows 9 to 0 hold 1, 3, 14, 68, 162, 162, 68, 14, 3 and 1 of them, where an element's row is the number of conditions in its basis; these match Wang's counts.
Element 0, in row 9, forces $A = 0$, and element 495, in row 0, has no conditions, so its problem is matrix multiplication over \Ftwo.
\textbf{(2)} Every list of conditions is equivalent to exactly one table element, so the table covers every class once.
\textbf{(3)} The table equals, element for element and in the same order, the list transcribed from our re-derivation of Wang's certificate, which keeps his classes and their order~\cite{wang2026automated}.
So the table's contents rest only on the definitions, and the transcription fixes only the numbering, making it easy to match our results against Wang's labeling.
\textbf{(4)} More conditions can only make the problem easier: lower bounds pass to fewer conditions, and upper bounds to more.
\textbf{(5)} Equivalent condition lists have the same bounds.
Facts (4) and (5) are Wang's Lemma~1~\cite{wang2026automated}, for the symmetries of matrix multiplication~\cite{degroote1978varieties,burichenko2016isotropy}, and they let a proof bound one element by looking up another.

\subsection{From the Table to Algorithms}

Layer 02 then ties the table's bounds to the algorithms of Section~\ref{sec:layer01}.

\textbf{(6)} A lower bound on the columns of valid decompositions is exactly a lower bound on the multiplications of programs that work on blocks, and the same holds for upper bounds, so nothing is lost by working with decompositions.
\textbf{(7)} A lower bound for element 495 is a lower bound for every program that works on blocks, because element 495 is the element with no conditions.

\begin{lstlisting}[style=lean]
theorem element_495_lower_bound_bounds_every_program (n : ℕ) :
    TableAtLeast ((table[495]!).map mask_matrix) n →
      RecursiveAlgorithmMultiplicationsAtLeast n
\end{lstlisting}

The theorem takes one input, $n$, and reads: if every decomposition valid for element 495 has at least $n$ columns, then every program that works on blocks has at least $n$ multiplications.
So a lower bound for matrix multiplication over \Ftwo is a lower bound for every recursive algorithm with integer constants.
It rests on the proof from Hopcroft and Kerr that an algorithm with integer constants, read modulo 2, is an algorithm over \Ftwo with no more multiplications~\cite{hopcroft1969minimizing}.
\textbf{(8)} A lower bound $n$ for element 495 and a valid decomposition with $n$ columns together make $n$ the fewest multiplications of any program that works on blocks (\lean{RecursiveAlgorithmMultiplicationsOptimalAt}).
Note that such a decomposition must be over the integers, because an algorithm for matrix multiplcation over \Ftwo does not necessarily imply an algorithm over the integers~\cite{hopcroft1969minimizing}. Lower bounds over \Ftwo transfer to the integers, but upper bound over \Ftwo do not.

\section{Sources and Contributions}
\label{sec:methods}
\label{sec:contributions}

\subsection{The Literature Used}
\label{sec:contributions:literature}

Throughout the project, the AI agents working on each upper or lower bound searched the literature to see whether their ideas had already been published, and the references below include what they found, with the software we used.
We have checked that each reference exists, but not that each contains exactly the idea we credit to it.

Wang's table, techniques and code are our basis~\cite{wang2026automated}, and Beuchert verified his certificate~\cite{beuchert2026independent}.
They build on the substitution method~\cite{pan1966methods,landsberg2017abelian,alexeev2011tensor,blaser2013fast} and on Hopcroft and Kerr~\cite{hopcroft1969minimizing,hopcroft1971minimizing}, whose reduction to bilinear algorithms also appears in~\cite{hopcroft1973duality} and extends to blocks through~\cite{amitsur1950minimal}.
Recursion~\cite{strassen1969gaussian,pan1984how}, Brent's equations~\cite{brent1970algorithms} and the symmetries of matrix multiplication~\cite{degroote1978varieties,brockett1978optimal,burichenko2016isotropy,marcus1959transformations} are classical.
The recent proofs giving a lower bound of 21 contain lemmas and certificates that we use or parallel~\cite{yang2026structural,wang2026lower21,tahir2026certificate}, including the rank-sum lemma and a duality for slices of defect one, from the research notes accompanying Yang et al.~\cite{yang2026structural}.
D'Ambrosio's work on a smaller format has related techniques~\cite{dambrosio2026exact}.
We also use standard facts from linear algebra and linear programming~\cite{steinitz1913bedingt,zhang2011matrix,farkas1902theorie,burgisser1997algebraic,landsberg2012tensors}.

The project began from~\cite{deza2023fast,heule2021new}.
Our searches used flip graphs~\cite{kauers2023flip,arai2024adaptive}, SAT solving~\cite{tseitin1983complexity,sinz2005towards,heule2012cube,soos2009extending}, branch and bound~\cite{land1960automatic}, symmetry breaking and orderly generation~\cite{crawford1996symmetry,read1978every,faradzev1978constructive}, tree-size estimates~\cite{knuth1975estimating}, screens~\cite{luks1997some,griesmer1960bound,macwilliams1963theorem,bland1977new,gray1953pulse}, and Laderman's algorithm gives the upper bound for matrix multiplication itself~\cite{laderman1976noncommutative}.
The software was HiGHS~\cite{huangfu2018parallelizing}, JuMP~\cite{lubin2023jump}, Julia~\cite{bezanson2017julia}, NumPy and SciPy~\cite{harris2020array,virtanen2020scipy}, and Lean with Mathlib~\cite{demoura2021lean4,mathlib2020}.

\subsection{Our Human Contributions}
\label{sec:contributions:authors}

We began by designing an exact search solver built on constraint propagation, and early on we pushed the AI agents to adapt its propagation techniques to individual elements of the table.
These exact searches told us each element's true value before we set out to prove it, so we were able to map the most promising routes to the lower bounds we needed for Theorem~\ref{thm:main}, and have the agents prove in Lean what we knew from the computation.

Additionally, the following glue theorem is ours.

\begin{lstlisting}[style=lean]
theorem decompositions_glue (C Ca Cb : List Bit3x3Matrix) (ma mb : ℕ)
    (h_meet : ∀ X, InSpan Ca X ∧ InSpan Cb X → InSpan C X)
    (ha : TableAtMost Ca ma) (hb : TableAtMost Cb mb) :
    TableAtMost C (ma + mb)
\end{lstlisting}

The theorem takes as inputs three lists of conditions, \lean{C}, \lean{Ca} and \lean{Cb}, and two numbers, \lean{ma} and \lean{mb}.
Its hypothesis \lean{h_meet} says that any condition implied by both \lean{Ca} and \lean{Cb} is also implied by \lean{C}.
Under \lean{h_meet}, and if \lean{Ca} has a decomposition with at most \lean{ma} columns and \lean{Cb} has a decomposition with at most \lean{mb}, then \lean{C} has one with at most \lean{ma + mb}.
In other words, when the inputs allowed by \lean{C} can be split between two easier problems, algorithms for those two problems add up to an algorithm for \lean{C}.
This theorem builds on ideas from Bläser~\cite{blaser2013fast} and Yang et al.~\cite{yang2026structural}.
Applied across the table, it gave 145 of our upper bounds.

\FloatBarrier
\section{Results}
\label{sec:results}

Table~\ref{tab:results} compares the bounds our Lean proof states with those Wang published~\cite{wang2026lower21}.
Our proof bounds all 496 elements of the table and determines 359 of them exactly, against 195 for Wang.
Its lower bounds exceed Wang's for 252 elements, by one for 245 and by two for 7, and match his for the rest.
All of its upper bounds come from our own decompositions, with the exception of the row representing matrix multiplcation without conditions, where the bound is due to Laderman~\cite{laderman1976noncommutative}.

Built from scratch, with Mathlib prebuilt, the proof's 3,521 Lean modules take 11.1 hours on one core, and about 3.7 hours with three modules built at a time, on an Apple M2 Pro with 12 cores and 32~GB of memory.

\begin{table}[!htb]
\centering

\footnotesize
\setlength{\tabcolsep}{3pt}
\begin{tabular}{c|c|>{\centering\arraybackslash}p{0.30\textwidth}|>{\centering\arraybackslash}p{0.27\textwidth}|cc|cc}
\hline
 & & & & \multicolumn{2}{c|}{exact} & & \\
row & $n$ & Wang & ours & Wang & ours & $+1$ & $+2$ \\
\hline
9 & 1 & 0 & 0 & 1 & 1 & 0 & 0 \\ \hline
8 & 3 & 3, 6, 9 & 3, 6, 9 & 3 & 3 & 0 & 0 \\ \hline
7 & 14 & 6$\times$2, 9$\times$6, 12$\times$5, 14 & 6$\times$2, 9$\times$6, 12$\times$5, 14 & 14 & 14 & 0 & 0 \\ \hline
6 & 68 & 9$\times$6, 12$\times$26, 13, 14$\times$11, [14,15]$\times$3, 15$\times$18, [15,16], 16, [17,18] & 9$\times$6, 12$\times$26, 13, 14$\times$11, 15$\times$21, 16$\times$2, 18 & 63 & 68 & 5 & 0 \\ \hline
5 & 162 & 11, 12$\times$10, [13,14]$\times$3, 14$\times$6, [14,15]$\times$13, 15$\times$60, [15,16]$\times$7, [15,17]$\times$8, 16$\times$5, [16,17]$\times$28, [16,18]$\times$15, [17,18]$\times$6 & 11, 12$\times$10, 14$\times$9, 15$\times$73, 16$\times$12, [16,17]$\times$5, [16,18], 17$\times$31, [17,18]$\times$18, 18$\times$2 & 82 & 138 & 68 & 5 \\ \hline
4 & 162 & 14$\times$3, [14,15], 15$\times$14, [15,16], 16, [16,17]$\times$49, [16,18]$\times$9, 17$\times$4, [17,18]$\times$57, [17,19]$\times$20, [17,20]$\times$3 & 14$\times$3, 15$\times$15, 16$\times$2, [16,17]$\times$8, 17$\times$41, [17,18]$\times$31, [17,19], [17,20], 18$\times$39, [18,19]$\times$17, [18,20]$\times$4 & 22 & 100 & 112 & 0 \\ \hline
3 & 68 & 15, [16,17]$\times$3, 17$\times$4, [17,18]$\times$13, [17,19]$\times$4, 18$\times$4, [18,19]$\times$16, [18,20]$\times$20, [18,21]$\times$3 & 15, 17$\times$7, 18$\times$16, [18,19]$\times$18, [18,21], 19$\times$2, [19,20]$\times$19, [19,21]$\times$4 & 9 & 26 & 45 & 0 \\ \hline
2 & 14 & 18, [18,19]$\times$4, [18,20], [19,20]$\times$3, [19,21]$\times$5 & 18, 19$\times$4, [19,20], 20$\times$2, [20,21]$\times$6 & 1 & 7 & 13 & 0 \\ \hline
1 & 3 & [19,21]$\times$3 & [20,21], 21$\times$2 & 0 & 2 & 1 & 2 \\ \hline
0 & 1 & [21,23] & [22,23] & 0 & 0 & 1 & 0 \\ \hline
total & 496 & & & 195 & 359 & 245 & 7 \\
\hline
\end{tabular}
\caption{Wang's table, by row, where $n$ is the number of elements in each row.
Each element is shown as its value when it is known exactly, or as an interval [lower bound, upper bound] otherwise, followed by their multiplicities ($\times$).
\emph{Wang}: the bounds Wang published~\cite{wang2026lower21}.
\emph{Ours}: the bounds our Lean proof states.
\emph{Exact}: the number of elements known exactly, for Wang and for us.
$+1$ and $+2$: the number of elements whose lower bound we raise by one and by two over Wang's.}
\label{tab:results}
\end{table}
\FloatBarrier

\section{Conclusion}
\label{sec:conclusion}

We proved in Lean that every $3\times3$ matrix multiplication algorithm with integer constants that can be applied recursively uses at least \lowerbound ~multiplications.
Applied recursively, such an algorithm cannot do better than $O(n^{\log_3 22}) \approx O(n^{2.814})$ multiplications.
It follows that no such algorithm can beat Strassen's $O(n^{\log_2 7}) \approx O(n^{2.807})$.
The proof builds on Wang's table: it bounds all 496 of its elements, determines 359 exactly, and raises lower bounds for 252.
Verifying the result requires auditing only a few short files of definitions and statements, and checking its axioms. 
Lean's kernel verifies the rest.

This proof leaves open the question of whether or not a 22-multiplication algorithm is possible for $3\times3$ matrix multiplication. We expect this gap to be closed imminently. We are refraining from publishing the code we used to produce the proof until one of the following two conditions is met: (1) we close the gap, or (2) this paper is accepted for publication.

\paragraph{Acknowledgments.} This research was undertaken, in part, thanks to funding from the Canada Research Chairs Program in Healthcare Analytics. We also acknowledge the support of the Natural Sciences and Engineering Research Council of Canada (NSERC RGPIN 2025-04995).

%
%
\bibliographystyle{splncs04}
\bibliography{bibliography}

\end{document}